\documentclass[12pt]{article}
\usepackage{amsmath,amssymb}
\usepackage{hyperref}

\def\beq{\begin{eqnarray}}
\def\eeq{\end{eqnarray}}

\newcommand{\Tr}{\,\mathrm{Tr}\,}            

\newcommand{\be}{\begin{equation}}
\newcommand{\ee}{\end{equation}}
\newcommand{\bea}{\begin{eqnarray}}
\newcommand{\eea}{\end{eqnarray}}
\newcommand{\bg}{\begin{gather}}
\newcommand{\eg}{\end{gather}}
\newcommand{\bseq}{\begin{subequations}}
\newcommand{\eseq}{\end{subequations}}

\renewcommand{\ln}{\mathop{\rm ln}\nolimits}

\def\tr{\hbox{Tr}}

\def\be{\begin{eqnarray}}
\def\ee{\end{eqnarray}}
\def\lb{\label}

\begin{document}

\title{\textbf{Entropy of random entangling surfaces }}
\vspace{1cm}
\author{ \textbf{
 Sergey N. Solodukhin$^\sharp$ }} 

\date{}
\maketitle

\begin{center}
  \hspace{-0mm}
  \emph{ Laboratoire de Math\'ematiques et Physique Th\'eorique, }\\
  \emph{Universit\'e Fran\c cois-Rabelais Tours, F\'ed\'eration Denis Poisson - CNRS, }\\
  \emph{Parc de Grandmont, 37200 Tours, France} \\
\end{center}

{\vspace{-11cm}
\begin{flushright}
\end{flushright}
\vspace{11cm}
}



\begin{abstract}
\noindent { We consider the situation when a globally defined four-dimensional field system is separated on two entangled  sub-systems by  a dynamical (random) 
two-dimensional surface.
The  reduced density matrix averaged over ensemble of random surfaces of fixed area and the corresponding average entropy are introduced.  
The average entanglement entropy is analyzed for a generic conformal field theory in four dimensions.
Two important particular cases are considered. In the first, both the intrinsic metric on the entangling surface and the spacetime metric  are fluctuating. 
An important example of this type is when the entangling surface is a  black hole horizon, the fluctuations of which cause necessarily the fluctuations in the spacetime geometry.
 In the second case, the spacetime is considered to be  fixed. 
The detail  analysis is carried out for the random entangling surfaces embedded in flat Minkowski spacetime.
In all cases the problem reduces to an effectively two-dimensional problem of random surfaces which can be treated by means of the well-known conformal methods.  Focusing on the logarithmic terms in the entropy we predict the appearance of a new $\ln\ln(A)$ term.}
\end{abstract}

\vskip 2 cm
\noindent
\rule{7.7 cm}{.5 pt}\\
\noindent 
\noindent
\noindent ~~~$^{\sharp}$ {\footnotesize e-mail: Sergey.Solodukhin@lmpt.univ-tours.fr}


\newpage

\section{ Introduction}
\setcounter{equation}0

Entanglement entropy \cite{BS}, \cite{Srednicki:1993im} has recently attracted much attention mainly due to its  geometrical nature
and the possible relevance to the entropy associated to the black hole and cosmological horizons. Much progress has been made in understanding
the entanglement entropy in the quantum field theories (for a review see \cite{Casini:2009sr}), in studying its holographic interpretation  \cite{Ryu-Takayanagi}
and in the promising applications to black holes (for a review see \cite{Solodukhin:2011gn}).

In the standard set-up the entanglement entropy is defined with respect to a given surface $\Sigma$ that can be any co-dimension two surface lying in the hypersurface 
of constant time. The intrinsic and extrinsic geometry of the surface is thus fixed. Altogether  with the geometry of the spacetime  this forms a fixed geometric background 
in the entropy calculation. 

In the present paper we consider a different, and perhaps more realistic, situation when the entangling surface is dynamical.
More precisely, we consider an ensemble of random entangling surfaces of fixed area and define a certain procedure of averaging over  ensemble. The average entropy which emerges in this new set-up is due to two factors: the quantum entanglement in the field system in question and the ``dynamics'' of the entangling surface itself.

This approach is well motivated in the case of the black hole horizons. Indeed, in the complete theory of Quantum Gravity the horizons are supposed to be dynamical
so that the horizon geometry  should fluctuate.  This may have many manifestations which however may be difficult to identify rigorously at the present stage 
of the theory. However, for the entropy, as we argue in this paper, the problem  of the horizon fluctuations can be  unambiguously treated already now.

Having in mind the possible applications to condensed matter, we remark that the dynamical surfaces appear in many systems studied theoretically and experimentally and 
it looks fairly  natural to associate with such surfaces  the entanglement entropy. This of course makes urgent  the necessity to address in the nearest future the question of experimental 
verification of the theoretical predictions. 

The paper is organized as follows. In section 2 we formulate the problem and define the average entanglement entropy. That the problem effectively reduces to a two-dimensional one
is illustrated in section 3 on a particular example of a product spacetime for which the usage of the heat kernel method is helpful. The general structure of the entanglement entropy
in a generic four-dimensional conformal field theory is analyzed in section 4. The average entropy is calculated in section 5 in the case when the fluctuations of the entangling surface are accompanied by the fluctuations of the spacetime, the quantum fluctuating horizon would be a typical example.  Entropy of random entangling surfaces embedded  in a fixed, for instance Minkowski,  spacetime is considered in section 6. We conclude with some remarks in section 7.

\section{ Formulation of the problem}
\setcounter{equation}0
In the standard set-up, in which the entanglement entropy is defined, one  assumes that the global quantum system is divided on two sub-systems by a surface $\Sigma$.
Suppose that this surface is closed and it is equipped with a metric $\gamma$.  The global system is  considered to be  in a vacuum state but after tracing over degrees of freedom residing
in one of the sub-systems one ends up with a reduced density matrix $\rho$.  
In the usual set-up the metric $\gamma$ is fixed and the reduced density matrix is a functional of $\gamma$. In fact, if the global system is defined in a curved space-time the reduced density matrix depends  on  the metric ($g$) of the spacetime in the vicinity of the entangling surface, on the intrinsic  metric $(\gamma$) and on the extrinsic curvature ($k$) of the surface, $\rho=\rho[\gamma, g, k]$. Respectively, the entanglement entropy, defined as
\be
S[\gamma, g, k]=-\Tr\rho\ln\rho=-(\alpha\partial_\alpha-1)\ln \Tr\rho^\alpha|_{\alpha=1}\, 
\lb{1}
\ee
depends on the intrinsic and extrinsic geometry of the surface and on the geometry of the spacetime\footnote{Additionally, the entropy may depend on the global quantum state of the field. Due to its complexity,  we do not consider this aspect in the paper.}.  It should be stressed that $\gamma$, $g$ and $k$ are not independent and are related by the so-called Gauss-Codazzi relations.

Let us suppose that this triple $(\gamma, g, k)$ is allowed   to  fluctuate (preserving of course the constraints imposed by the Gauss-Codazzi relations) in such a way that the area $A$ of $\Sigma$,  the genus $h$, defined as $(1-h)=\frac{1}{8\pi}\int_\Sigma R_\Sigma$, and, possibly, some other integral conditions on the geometry of $\Sigma$  are fixed.
Thus we have to deal with a statistical ensemble of triples  on $\Sigma$.
Naturally, we have to define an average over this ensemble,
\be
<\Tr\rho^\alpha>=\frac{\int {\cal D}\gamma\, {\cal D}g\, {\cal D}k\, \delta(A-\int_\Sigma\sqrt{\gamma})\, \Tr\rho^\alpha[\gamma]}{\int {\cal D}\gamma\, {\cal D}g\, {\cal D}k\, \delta(A-\int_\Sigma\sqrt{\gamma})\, \Tr\rho[\gamma]}\, .
\lb{2}
\ee
The normalization  in (\ref{2}) is such  that $<\Tr \rho>=1$.
The average entropy is then defined in a close analogy with (\ref{1}) as follows,
\be
<S>=-<\Tr\rho\ln\rho >=-(\alpha\partial_\alpha-1)\ln <\Tr\rho^\alpha>|_{\alpha=1}\, .
\lb{3}
\ee
This quantity incorporates two effects: the entanglement between two sub-systems for a given triple  and the averaging over statistical ensemble of triples.

\bigskip

In this paper we study two problems: 

\medskip

\noindent i) for a fixed surface $\Sigma$ embedded in a given spacetime we analyze  the dependence of the entanglement entropy (\ref{1}) on the triple $(\gamma, g, k)$;  

\medskip

\noindent ii)  we identify  the dependence  on the area $A$ of the average entropy (\ref{3}).  

\bigskip

Depending on the way the entangling surface is defined the two situations are possible: 

\medskip

\noindent 1) The bulk spacetime metric $(g)$ fluctuates together with the intrinsic metric $(\gamma)$ of the surface. This happens when  the entangling surface is defined essentially by the structure of the  spacetime. So that a deformation of the geometry on the surface is necessarily  accompanied by a certain deformation in the spacetime metric and vise versa. 
The most interesting example of this type is the fluctuating horizon. Indeed, provided the bulk spacetime satisfies the Einstein equations and the asymptotic charges (mass, electric charge and  angular rotation) are specified, the geometry (and, in four dimensions, the  topology) of the horizon is completely fixed and can not fluctuate. The fluctuations of the horizon may however happen in a quantum description when the spacetime in a small region near the horizon randomly deviates from the classical and the geometry of the horizon itself fluctuates.  Classical horizon is a minimal surface for which the extrinsic curvature vanishes, $k=0$
The fluctuating horizon is supposed to remain  a minimal surface so that all components of the extrinsic curvature of the horizon would  still vanish.  The other relevant situation is when $\Sigma $ is the neck of a wormhole (the extrinsic curvature of $\Sigma$ vanishes in this case too). The fluctuations of the geometry of $\Sigma$ are thus  accompanied by the fluctuations in spacetime in a small region close to $\Sigma$.

As we will see the situation when both the surface geometry and the bulk metric fluctuate,  is the most tractable case when the procedure of computation of the average entropy (\ref{3}) can be completely carried out. This is due to the fact that the conformal fluctuations of the two-dimensional (2d)  entangling surface may be
enlarged to conformal fluctuations of the bulk spacetime. The conformal symmetry then helps to proceed with the entropy computation.   

\medskip

\noindent 2) The bulk spacetime  metric is fixed and  non-fluctuating. For example, the spacetime is  Minkowski space. The geometry of  the entangling surface, however, may fluctuate that can be easily visualized as random deformations of the shape of  the surface embedded in Minkowski spacetime. Rather surprisingly, this
seemingly simpler situation is  more difficult to approach. Even though the fluctuations of the surface still can be represented by a fluctuating conformal factor there is no
conformal symmetry in this case since the background spacetime remains  fixed and non-fluctuating.

\section{ Reduction to an effective two-dimensional problem: an illustration}
\setcounter{equation}0
The calculation of the entropy (\ref{1}) in the standard set-up can be carried out by using the so-called conical singularity method, in some detail this method is explained in  
\cite{Solodukhin:2011gn}. It consists in introducing a small conical singularity with angle deficit $\delta=2\pi (1-\alpha)$ in the two-dimensional sub-space orthogonal
to the entangling surface $\Sigma$ so that locally, in a small vicinity of $\Sigma$, the space-time ${\cal M}_\alpha$ looks like a direct product $C_{2,\alpha}\times \Sigma$ of two-dimensional conical space $C_{2,\alpha}$ and surface $\Sigma$.
If the field in question is bosonic  and is described by a field operator $\hat{D}$ then one has a representation
\be
-\ln\Tr\rho^\alpha =W(\alpha)=\frac{1}{2} \ln\det \hat{D}_{{\cal M}_\alpha}\, 
\lb{4}
\ee 
in terms of determinant of operator $\hat D$ on the conical space-time ${\cal M}_\alpha$. The standard way to calculate (\ref{4}) is to use the heat kernel,
\be
W(\alpha)=-\frac{1}{2}\int_{\epsilon^2}^\infty \frac{ds}{s}\Tr K_{{\cal M}_\alpha}(s)\, ,
\lb{5}
\ee
where $K_{{\cal M}_\alpha}=e^{-s \hat{D}}$ is the heat kernel of operator $\hat D$.
In general the conical space ${\cal M}_\alpha$ may be rather complicated. 
However,  in order to illustrate our main idea we consider the simplest case when it is a direct product not only locally but also globally, ${\cal M}_\alpha =C_{2,\alpha}\times \Sigma$.
Then the analysis is quite simple and can be carried out with the help of the heat kernel. Indeed, in this case  the heat kernel is the product 
\be
\Tr K_{{\cal M}_\alpha}(s)=\Tr K_{C_{2,\alpha}}(s) \cdot \Tr K_\Sigma (s)\, .
\lb{6}
\ee
This property follows from the fact that on a product of two spaces the differential operator is sum of operators acting on each space, $\hat{D}_{{\cal M}_\alpha}=\hat{D}_{C_{2,\alpha}}+\hat{D}_\Sigma$.
Suppose for simplicity that operator $\hat{D}$ is (minus) the Laplace operator, $\hat{D}=-\Delta$, describes a minimally coupled massless scalar field. Then $\hat{D}_\Sigma=-\Delta_\Sigma$ is the Laplace operator on 2d surface $\Sigma$.  The heat kernel on the two-dimensional cone is known explicitly (see, for instance, \cite{Dowker:1977zj} and \cite{Fursaev:1993qk})\footnote{In this paper only massless quantum fields will be considered.
Generalization to massive fields, at least in Minkowski spacetime, is rather straightforward, see for instance \cite{Calabrese:2004eu}.}
\be
\Tr K_{C_{2,\alpha}}(s)=\alpha\Tr K_{R_2}(s)+d(\alpha)\, , \ \ d(\alpha)=\frac{(1-\alpha^2)}{12\alpha}\, ,
\lb{7}
\ee
where $\Tr K_{R_2}(s)=\frac{V(R_2)}{4\pi s}$ is the heat kernel on two-dimensional plane $R_2$ obtained from $C_{2,\alpha}$ by setting $\alpha=1$, $V(R_2)$ is volume of $R_2$.
Combining (\ref{6}) and (\ref{7}) we arrive at the following form for the effective action (\ref{5}) 
\be
W(\alpha)=\alpha \, W(1)+\frac{d(\alpha)}{2}\ln\det(- \Delta_\Sigma)\, ,
\lb{8}
\ee
where we used that 
\be
\ln\det (-\Delta_\Sigma)=-\int_{\epsilon^2}^\infty \frac{ds}{s} \Tr K_\Sigma (s)\, .
\lb{9}
\ee
Thus, the trace of $\alpha$-th power of the reduced density matrix, up to some irrelevant factor, is represented as a functional integral over  $d(\alpha)$ (fictitious) scalar fields $X^a$ defined on the 2d surface $\Sigma$,
\be
&&\Tr\rho^\alpha=e^{-\alpha W(1)}(\det(-\Delta_\Sigma))^{-d(\alpha)/2}\lb{10} \\
&&=e^{-\alpha W(1)} \int {\cal D} X e^{-S(X,\gamma)}\, , \ \ S(X,\gamma)=\frac{1}{2}\int_\Sigma \sqrt{\gamma}\gamma^{ij}\partial_i X^a \partial_j X^a \, .\nonumber
\ee
The functional integral in two dimensions is given explicitly
\be
\int {\cal D} X e^{-S(X,\gamma)}=e^{-d(\alpha)W_\Sigma(\gamma)}\, ,
\lb{11}
\ee
where we introduced (see, for instance, \cite{Frolov:1996hd} and references therein) 
\be
W_\Sigma(\gamma) =-\frac{A(\gamma)}{8\pi\epsilon^2}-\frac{(1-h)}{3}\ln \epsilon^{-1} +\frac{1}{96\pi}\int_\Sigma R_\Sigma \frac{1}{\Delta_\Sigma} R_\Sigma\, 
\lb{12}
\ee
and $(1-h)=\frac{1}{8\pi}\int_\Sigma R_\Sigma$.
Applying (\ref{1}) and using that $d(\alpha)=\frac{1}{6}(1-\alpha)+O(1-\alpha)^2$, one finds 
\be
S(\gamma)=-\frac{1}{6}W_\Sigma(\gamma)\, 
\lb{13}
\ee
for the entanglement entropy.
For a  typical geometry  characterized by just one scale parameter $a$, so that the area $A(\gamma)\sim a^2$,  we have  
\be
S(\gamma)=\frac{A(\gamma)}{48\pi\epsilon^2}+\frac{(1-h)}{36}\ln \frac{A(\gamma)}{\epsilon^2}\, 
\lb{14}
\ee
for the entropy. A finite $\ln A(\Sigma)$ term in (\ref{14}) comes from the non-local Polyakov term in (\ref{12}) after the rescaling of the metric, $\gamma\rightarrow a \gamma$, see
\cite{Frolov:1996hd} for more detail. 
 The parameter $\epsilon$ is an UV cut-off introduced in (\ref{9}) to regularize the integral over the proper time. Later in the paper we will use an UV regularization with a parameter  $\epsilon$ normalized in such a way that
the area term in the entropy will always take the same form as in (\ref{14}).

Suppose now that the geometry of the surface $\Sigma$ is fluctuating provided the direct product structure $R_2\times \Sigma$ of the spacetime is preserved. 
Then the quantity which we want to compute is the average (\ref{2}) over ensemble of random geometries. The extrinsic curvature of $\Sigma$ in the direct product vanishes. 
On the other hand, the fluctuations of the bulk metric just reduce to that of the surface $\Sigma$. Thus, the functional integral in (\ref{2}) reduces to the integral over random
intrinsic metrics $\gamma$ only. The calculation of the average of (\ref{10}) over intrinsic metrics is directly related to the problem already studied in the literature, see
 \cite{Distler:1988jt} and \cite{Zamolodchikov:1982vx}, and can be performed   by means of the methods of two-dimensional conformal field theory. This old calculation is helpful in our case. In the paragraph below we very briefly summarize the results  of  \cite{Distler:1988jt} useful for our purposes.

Let us introduce the partial partition function ${\cal Z}_d(A(\Sigma))$ as follows
\be
\int {\cal D} \gamma {\cal D} X e^{-S(X,\gamma)}\delta\left(\int_\Sigma \sqrt{\gamma}-A(\Sigma)\right)= e^{-dA(\Sigma)\over 8\pi\epsilon^2}{\cal Z}_d(A(\Sigma))\, ,
\lb{14-1}
\ee
where the functional integral is taken over $d$ scalar fields $X$,  ${\cal D} \gamma$ is the DDK measure which includes the gauge fixing and the ghost contribution.
It has the following scaling property \cite{Distler:1988jt}
\be
&&{\cal Z}_d(A(\Sigma))=e^{(Q(1-h)/q-1)\rho}~{\cal Z}_d(A(\Sigma)e^{-\rho})\, , \nonumber \\
&&Q=\sqrt{\frac{25-d}{3}}\, , \ \ q=-\frac{Q}{2}+\frac{1}{2}\sqrt{Q^2-8}\, .
\lb{14-2}
\ee
One finds that \cite{Distler:1988jt}
\be
{\cal Z}_d(A)=C A^{(1-h)Q/q-1}\, ,
\lb{14-3}
\ee
where $C$ is an irrelevant constant.
With the normalization as in (\ref{2}) we then find that the average trace of the reduced density matrix  can be presented as follows
\be
&&<\Tr \rho^\alpha >= e^{-\frac{d(\alpha)A}{8\pi\epsilon^2}} {\cal Z}_{d(\alpha)}(A(\Sigma))/{\cal Z}_{d=0}(A(\Sigma)) \nonumber \\
&&=e^{-\frac{d(\alpha)A}{8\pi\epsilon^2}}\left(\frac{A}{\epsilon^2}\right)^{\Gamma(\alpha)-\Gamma(1)}\, ,
\lb{15}
\ee
where $\Gamma(\alpha)$ is the known quantity (called the string  susceptibility) \cite{Distler:1988jt}
\be
\Gamma(\alpha)=\frac{1}{12}(1-h)[d(\alpha)-25-\sqrt{(25-d(\alpha))(1-d(\alpha))}]+2\, 
\lb{16}
\ee
and $d(\alpha)$ is given in (\ref{7}).  The formulas (\ref{14-1})-(\ref{16}) are valid if $d<1$. In our case $d$ is close to zero (since $\alpha\sim 1$) 
and these formulas are perfectly valid.

For small $(1-\alpha)$ one has that $d(\alpha)\simeq \frac{1}{6}(1-\alpha)$ and hence $\Gamma (\alpha)-\Gamma(1)\simeq \frac{1}{20}(1-h)(1-\alpha)$. So that one finds 
\be
<S>=\frac{A(\Sigma)}{48\pi\epsilon^2}+\frac{(1-h)}{20}\ln \frac{A(\Sigma)}{\epsilon^2}\, 
\lb{17}
\ee
for the average entropy (\ref{3}).
Comparison with the expression (\ref{14}) shows that the procedure of averaging over ensemble of random surfaces changes  the logarithmic term in the entropy.

We note that in this example the bulk metric fluctuates together with the surface $\Sigma$. Moreover, the extrinsic curvature of $\Sigma$ in the direct product $R_2\times \Sigma$ vanishes.
Thus this example may be relevant to the entropy of a fluctuating horizon in a certain extreme limit \cite{Mann:1997hm} of the black hole geometry.

\section{A generic 4d CFT and the entropy of arbitrary entangling surface }
\setcounter{equation}0

Consider now a generic 4d conformal field theory, characterized by the conformal anomalies of type A and B. On a curved background (equipped with metric $g_{\mu\nu}$)
with a singular surface $\Sigma$ (equipped with 2d metric $\gamma_{ij}$) this theory  is described by the quantum effective action which has the bulk and boundary components. To first order in $(1-\alpha)$ one has that
\be
W=\alpha W_{bulk}+(1-\alpha){\cal W}_\Sigma 
\lb{18}
\ee
The bulk part of the action has a standard decomposition in terms of the UV cut-off $\epsilon$ 
\be
W_{bulk}=\int_M \left({a_4\over \epsilon^4}+{a_1\over \epsilon^2}-a_2\ln\epsilon\right) +w(g)\,  .
\lb{19}
\ee
Under the conformal transformations of the bulk metric,  $g\rightarrow e^{2\sigma(x)} g$, the UV finite part $w(g)$ transforms as follows
\be
w( e^{2\sigma} g) =w(g)+\int_M a_2\sigma(x)+O(\sigma^2)\, .
\lb{20}
\ee
The integral of $a_2$ is conformal invariant.  It
 is the so-called conformal anomaly. Quite generically, the anomaly may contain the topological invariants (anomaly of type A) or powers of the Weyl tensor (anomaly of type B). In four dimensions one has that (see \cite{Duff:1993wm} for a review)
\be 
&&a_2=A E_{(4)}+B I_{(4)}\, ,\nonumber \\
&&E_{(4)}={1\over 64}
(R_{\alpha\beta\mu\nu}R^{\alpha\beta\mu\nu}-4R_{\mu\nu}R^{\mu\nu}+R^2)\, ,
\nonumber \\ &&I_{(4)}=-{1\over
64}(R_{\alpha\beta\mu\nu}R^{\alpha\beta\mu\nu}-2R_{\mu\nu}R^{\mu\nu}+{1\over
3} R^2)\, .
\label{21} 
\ee 
The transformation (\ref{20}) can be ``integrated'' to reproduce that part in the effective action $w(g)$ that generates the conformal anomaly, of course the integration is only up to an ``integration constant'', a conformally invariant part in the effective action.

Similarly, the surface term in the effective action (\ref{18}) is decomposed on the UV divergent and UV finite parts,
\be
{\cal W}_\Sigma=-\int_\Sigma\left({N\over 48\pi\epsilon^2}+s_0\ln\epsilon\right) -s(g,\gamma)\, ,
\lb{22}
\ee 
where $N=N(A,B)$ is the effective number of fields in the theory, the fermions are counted with the weight $1/2$.
Under the conformal (Weyl) transformations of the  bulk metric and of the surface metric, $g\rightarrow e^{2\sigma} g$, $\gamma\rightarrow e^{2\sigma} \gamma$ the UV finite part $s(g, \gamma)$ transforms as
\be
s(e^{2\sigma} g,e^{2\sigma}\gamma)=s(g,\gamma)-\int_\Sigma s_0 \sigma +O(\sigma^2)\, .
\lb{23}
\ee 
The term $s_0$ is the surface part of the conformal anomaly. As its bulk counterpart it consists of anomalies of type A and B, see ref. \cite{Solodukhin:2008dh} (and \cite{Ryu-Takayanagi} for the black hole case  when the extrinsic curvature does not contribute): 
\be
&&s_0=A s_A+B s_B \,, \nonumber \\
 &&s_A=
{\pi\over 8}R_\Sigma \, , \ \ s_B= -{\pi\over 8}K_\Sigma\, ,
\label{24} 
\ee 
where $R_\Sigma$ is the intrinsic curvature of surface $\Sigma$ and we introduced the quantity \cite{Solodukhin:2008dh}
\be
K_\Sigma=R_{ijij}-R_{ii}+{1\over
3}R -(\tr k^2-{1\over 2} k_i k_i )\, ,
\label{25} 
\ee 
where $R_{ijij}=R_{\alpha\beta\mu\nu} n^\alpha_i n^\beta_j n^\mu_i n^\nu_j$, $R_{ii}=R_{\alpha\beta} n^\alpha_i n^\beta_i$, and $n_i^\mu$, $i=1,2$ are two vectors normal to $\Sigma$. The quantity (\ref{25}) 
is determined by both the intrinsic and extrinsic geometry of the entangling surface $\Sigma$.
Other forms of the surface anomaly can be obtained by using the 
Gauss-Codazzi equation 
\be
R=R_\Sigma +2R_{ii}-R_{ijij}-k^ik^i+\tr
k^2\, .
\label{26}
\ee
Under the conformal transformations the intrinsic curvature $R_\Sigma$ and the quantity $K_\Sigma$ transform as follows
\be
&&R_\Sigma(e^{2\sigma} \gamma)=e^{-2\sigma}(R_\Sigma(\gamma)-2\Delta_\Sigma\sigma)) \nonumber \\
&&K_\Sigma(e^{2\sigma} g,e^{2\sigma}\gamma))=e^{-2\sigma}K_\Sigma(g,\gamma)\, .
\lb{27}
\ee
The surface integral of $s_A$ is topological invariant, the Euler number of the surface. On the other hand, the surface integral of (\ref{25})
\be
{\cal K}=\frac{1}{8\pi} \int_\Sigma K_\Sigma=\frac{1}{8\pi} \int_\Sigma \left(R_{ijij}-R_{ii}+{1\over
3}R -(\tr k^2-{1\over 2} k_i k_i )\right)\, 
\label{26-1} 
\ee 
 is conformal invariant. 

With the help of (\ref{27}) one can integrate (\ref{23}) and obtain the non-local  surface action
\be
s(\gamma, k, g)={A\pi\over 32}\int_\Sigma R_\Sigma {1\over \Delta_\Sigma} R_\Sigma -{B\pi\over 16}\int_\Sigma K_\Sigma {1\over \Delta_\Sigma} R_\Sigma +s_{conf}(\gamma, k, g)
\, .
\lb{28}
\ee
The last  term, a ``constant of integration'',  is conformal invariant. It can be represented in the form
\be
s_{conf}(\gamma, k, g)=C_1\int_\Sigma K_\Sigma+ C_2\int_\Sigma K_\Sigma {1\over \Delta_\Sigma} K_\Sigma+C_3 \int_\Sigma K_\Sigma {1\over \Delta_\Sigma} K_\Sigma {1\over \Delta_\Sigma} K_\Sigma+ ..\, ,
\lb{29}
\ee
where $C_1, \ C_2,\ C_3, ..$ are some undetermined constants. We neglect in (\ref{28}) the possible topological  term
$
\mu\int_\Sigma R_\Sigma
$. It is interesting to note that it is impossible to write down any  local or non-local conformal invariants in terms of only the intrinsic metric
on two-dimensional surface. On the other hand, by including the extrinsic geometry of the surface in the consideration one is able to write an infinite family of conformal invariants
as in (\ref{29}).

Calculating the entanglement entropy for an entangling surface $\Sigma$, provided the bulk metric $g$,  the surface metric $\gamma$ and the extrinsic curvature $k$ are fixed, we come to the formula
\be
&&S(\gamma, k, g)=-{\cal W}_\Sigma(\gamma, k, g)\nonumber \\
&&={N\, A(\gamma)\over 48\pi\epsilon^2}+s_0\ln\epsilon+s(\gamma, k, g)\, ,
\lb{30}
\ee 
where $s_0$ is given by  (\ref{24}) and  $s(\gamma, k, g)$ is given by (\ref{28}).  Notice, that we did not yet make any restrictions neither on the extrinsic geometry nor on the geometry of the bulk spacetime. This is the conformal symmetry that helped us to obtain this general result valid for the entanglement entropy of arbitrary entangling surface.

\bigskip

\noindent Consider  two limiting cases.

\medskip

\noindent i). Suppose that invariant $K_\Sigma=0$. In this case the log term in the  entropy  (\ref{30}), (\ref{28}) is due the anomaly of type A only and it depends only on the intrinsic geometry of surface $\Sigma$.  For a typical geometry characterized by scale $a$ one finds
\be
S(\gamma)={N\, A(\gamma)\over 48\pi\epsilon^2}-{A\pi^2(1-h)\over 2}\ln{A(\gamma)\over \epsilon^2}\, ,
\lb{31}
\ee
where $A(\gamma)\sim a^2$.

\medskip

\noindent ii).  Suppose that the intrinsic curvature of the surface is vanishing, $R_\Sigma=0$. It is the case if for example the surface $\Sigma$ is torus (or cylinder in non-compact case). Then  the A-anomaly term in (\ref{28}) vanishes. The B-anomaly term however is non-vanishing  and is equal to
\be
s(\gamma, k, g)= -{B\pi\over 16}\int_\Sigma K_\Sigma \psi_\Sigma+s_{conf}(\gamma, k, g)
\, ,
\lb{32}
\ee
where $\psi_\Sigma$ is a harmonic function on $\Sigma$, $\Delta_\Sigma\psi_\Sigma=R_\Sigma=0$.

\bigskip

A couple of remarks are in order. 

\noindent 1. Comparing the logarithmic terms in expressions (\ref{31}) and (\ref{14}) we note that they come with an opposite sign. It is explained by the observation that for a
product metric $R_2\times S_2$ the extrinsic curvature  as well as the quantities $R_{ijij}$ and $R_{ii}$ vanish  so that one has that $K_\Sigma=R_\Sigma/3$. Taking that for a scalar field $A=B/3$ one finds that the $A$ and $B$ terms in (\ref{28}) combine in a negative quantity. Also, we note that the calculation in Section 3 was made for a non-conformal scalar field.

\noindent 2. If the surface $\Sigma$ is embedded in flat spacetime then from the Gauss-Codazzi equation (\ref{26}) we find that the intrinsic curvature of the surface is directly related to the 
extrinsic curvature via relation $R_\Sigma=k^i k^i-\tr k^2$.  So that one has that $K_\Sigma=1/2(R_\Sigma-\tr k^2)$.  It is curious to note that the local part of the surface action (\ref{22}) and/or of the entropy (\ref{30}) then coincides with the action of the so-called rigid string introduced by Polyakov in \cite{Polyakov:1986cs}.

\section{ Entropy of two-dimensional random surface in fluctuating spacetime}
\lb{fluct-metric}
\setcounter{equation}0
As we have seen in the previous section, the conformal symmetry plays an important role in the identifying the structure of the surface action and of the entropy. The symmetry is generated by  the  simultaneous transformations of the bulk spacetime metric ($g$) and of the surface metric $(\gamma$),
\be
&&g(r,t, x)\rightarrow e^{2\sigma(r, t, x)} g(r,t,  x) \, , \, \, \gamma(x)\rightarrow e^{2\sigma(x)} \gamma(x)\nonumber \\
&&\sigma(r, t, x)=(\sigma(x)+\sigma_1(x) r+..)+O(t) \, ,
\lb{33}
\ee
where $(x)$ stands for the intrinsic coordinates on surface $\Sigma$ and $r$ is a radial coordinate orthogonal to $\Sigma$ on the hypersurface $t=0$. The surface $\Sigma$ is thus defined by the conditions: $t=0, r=0$.
The surface action defined in the previous section is presented as a sum of a term invariant under the transformation (\ref{33}) and the anomaly. Neglecting the conformally invariant part we have that
\be
&&W_\Sigma=(1-\alpha) {\cal W}_\Sigma \, , \nonumber \\
&&{\cal W}_\Sigma=-{N A(\Sigma)\over 48\pi\epsilon^2}-{A\pi\over 32}\int_\Sigma R_\Sigma {1\over \Delta_\Sigma} R_\Sigma +{B\pi\over 16}\int_\Sigma K_\Sigma {1\over \Delta_\Sigma} R_\Sigma \, .
\lb{34}
\ee
As we have seen in (\ref{10}) the scalar $K_\Sigma$ transforms homogeneously under (\ref{33}) so that the surface integral $\int_\Sigma K_\Sigma$ is invariant under (\ref{33}).

\bigskip

Consider now a statistical  ensemble of triples $(\gamma, g, k)$ such that 

\medskip

\noindent 1) the area of the surface $\Sigma$ is fixed, $A(\Sigma)=\int_\Sigma \sqrt{\gamma}$;

\medskip

\noindent 2) the topological type $h$ of the surface is fixed, $(1-h)=\frac{1}{8\pi}\int_\Sigma R_\Sigma$;

\medskip

\noindent  3) the conformal invariant ${\cal K}=\frac{1}{8\pi} \int_\Sigma K_\Sigma$ of the surface $\Sigma$ is fixed.

\medskip

 In  a small vicinity of $\Sigma$ the fluctuations of the bulk metric $g$ can be decomposed onto the fluctuations of the 2d metric $\gamma$ and the fluctuations $g_{\bot}$ in the directions orthogonal to the surface, so that the integration measure in the statistical ensemble of triples  factorizes
\be
{\cal D} \gamma  {\cal D} g_{\bot} {\cal D} k \, \delta(A(\Sigma)-\int_\Sigma \sqrt{\gamma})\, \delta({\cal K}-\frac{1}{8\pi} \int_\Sigma K_\Sigma)\, ,
\lb{35}
\ee
where ${\cal D} \gamma $ is the DDK measure \cite{Distler:1988jt}. 
We remind the reader that we are interested in that part of the fluctuations which are governed by the surface action (\ref{34}). In this action the ``orthogonal'' fluctuations are important for the dynamics governed by the conformally invariant part of the surface action (not shown in (\ref{34})). The averaging over  ensemble thus gives
\be
<\Tr\rho^\alpha>\sim \int {\cal D}\gamma e^{-(1-\alpha){\cal W}_\Sigma} \int {\cal D}g_{\bot} \, {\cal D} k \, e^{-(1-\alpha){\cal W}_{conf}}\, ,
\lb{36}
\ee
where in the both integrals we,  for compactness, skip the delta-functions.  
Moreover, if the surface $\Sigma$ is specified by a certain condition which specifies the way the surface is embedded in the larger spacetime, this condition is formulated in terms of the components of the metric  in the ''orthogonal'' direction. This condition thus  
constrains the functional integral over $g_{\bot}$ in (\ref{36}) and does not affect the integral over intrinsic metrics $\gamma$. 

We note that the integral over $g_{\bot}$  and $k$ is scale invariant while the dependence on the scale comes from the integral over $\gamma$.
Let us focus on the integral over $\gamma$ in (\ref{36}),
\be
Z(\alpha, A(\Sigma))=\int {\cal D}\gamma \, \delta(A(\Sigma)-{\int}_\Sigma \sqrt{\gamma}) e^{-(1-\alpha){\cal W}_\Sigma}\, .
\lb{37}
\ee
 In the surface action (\ref{34}) the role of A and B terms is different. The A-term makes a contribution to the central charge.
This central charge is the same as the one produced by $d(\alpha)$ of fictitious 2d scalar fields, $d(\alpha)=-3A\pi^2(1-\alpha)+O(1-\alpha)^2$.  On the other hand, the B-term   just gives a  coupling of the conformal factor $\sigma$ of the metric, $\gamma=e^{2\sigma}\delta$, to an external operator $K_\Sigma$. This coupling does not contribute to the central charge but affects the scaling properties of the action $W_\Sigma$. The scaling transformation of (\ref{37}) follows from the scaling properties of the DDK measure and  one has that
(with the normalization as in (\ref{2}))
\be
&&Z(\alpha, A(\Sigma))=e^{-\frac{(1-\alpha) N\, A(\Sigma)}{48\pi\epsilon^2}} A(\Sigma)^{(\bar{\Gamma}(\alpha)-\bar{\Gamma}(1))}\, , \lb{38}  \\
&&\bar{\Gamma}(\alpha)={\Gamma}(\alpha)+(1-\alpha){B\pi^2{\cal K}\over 2}\, , \nonumber 
\ee
where ${\Gamma}(\alpha)$ is given by (\ref{16}) and $d(\alpha)=-3A\pi^2(1-\alpha)$. The quantity (\ref{38}) is the scale dependent part of the integral (\ref{36})
so that the scale dependent part in the  entropy is
\be
<S>=\frac{N\, A(\Sigma)}{48\pi\epsilon^2}-\left(\frac{9}{10}A\pi^2(1-h)-\frac{1}{2}B\pi^2 {\cal K}\right)\ln A(\Sigma)\, .
\lb{39}
\ee
This equation is one of  our main results. It is supposed to be rather general since all the details on how  the class of fluctuating entangling surfaces  is specified
are contained in the integral over $g_{\bot}$ and do not affect the integral over $\gamma$. Thus, the integral over $\gamma$ and  the entropy (\ref{39}) are  universal.

The application of (\ref{39}) to the case when $\Sigma$ is a fluctuating horizon is of course the most interesting. The fluctuations of horizons naturally occur in Quantum Gravity 
and we claim that our result (\ref{39}) is the entropy of a (black hole or cosmological) horizon   in the complete theory of Quantum Gravity, provided the contributions of all fields present in the theory (including the gravitons) are taken into account in the anomaly coefficients $A$ and $B$. The horizon surface is topologically sphere, so that $h=0$.
The invariant $\cal K$ however is not universal and depends on the type of the black hole. For the Kerr-Newman black hole characterized by mass $m$, electric charge $q$ and
rotation parameter  $a$ we find (using the useful expressions for the surface integrals obtained in \cite{Mann:1996bi}) that 
\be
{\cal K}=1-\frac{q^2}{3r_+^2}\left(1+\frac{3}{2}(\frac{r_+^2+a^2}{ar_+})\tan (\frac{a}{r_+})\right)\, ,
\lb{39-1}
\ee
where $r_+=m+\sqrt{m^2-q^2-a^2}$.
For a large class of uncharged black holes ($q=0$),  the Schwarzschild black hole, the Kerr black hole including the extreme Kerr black hole, one has that ${\cal K}=1$
so that the logarithmic term in the entropy (\ref{39}) is proportional to $(\frac{9}{5}A-B)$. This is different from the log  term in the entropy of the classical   horizon that is proportional to $(A-B)$, see \cite{Solodukhin:2011gn}. Thus, for a CFT with $A=B$ the log term in the entropy of a classical Kerr black hole would vanish  while it would reappear in the entropy of a quantum (fluctuating) horizon. The other interesting limit corresponds to an extreme charged black hole for which $h=0$ and ${\cal K}=0$. The log term in the entropy of classical horizon in this case depends only on the anomaly of type A (see \cite{Solodukhin:2011gn}). This is still true for the entropy of a quantum (fluctuating) extreme horizon (provided the fluctuations do not change the value of $\cal K$-invariant  of the horizon) although the exact pre-factor in the log term changes.

\section{ Random entangling surfaces in Minkowski spacetime}
\setcounter{equation}0
Let us discuss now the situation when the  entangling surface $\Sigma$ is embedded in a  spacetime whose geometry is fixed and non-fluctuating. 
For simplicity and concreteness let us consider the case when the spacetime in question
is the 4D Minkowski spacetime, the surface $\Sigma$ lies in the hypersurface  of constant time $t$,  a flat Euclidean space $R^3$. One of the vectors normal   to $\Sigma$,  say $n^2$, is time-like and thus is  orthogonal to the $t=const$ hypersurface. The corresponding extrinsic curvature $k^2$  identically vanishes.  The other normal vector, $n^1$, lies entirely in the $t=const$  surface $R^3$ and has a non-vanishing extrinsic curvature $k^1=k$ which is  determined by the way $\Sigma$ is  embedding  in the 3d space $R^3$.
Since  the  bulk curvature vanishes, we have the following expressions for the Ricci scalar and the scalar $K_\Sigma$ of the entangling surface
\be R_\Sigma=(\tr k)^2-\tr k^2\, , \ \ K_\Sigma=\frac{1}{2} (\tr k)^2-\tr
k^2\, .
\label{40}
\ee
As soon as  the bulk geometry is fixed, we are allowed only to conformally transform the intrinsic metric $\gamma$  of the surface $\Sigma$ and not to touch the bulk metric $g$ . The quantity ${\cal K}$ is not invariant under the conformal transformations of only $\gamma$. Thus the (conformal) fluctuations of the intrinsic metric will cause the quantity ${\cal K}$ to fluctuate.  This indicates that the  averaging procedure is more delicate in this case and should take into account the fluctuating extrinsic curvature. 
The functional integration in (\ref{2}) thus reduces to the integration over $\gamma$ and $k$. 
One, a rather standard, way to treat the fluctuations of $\gamma$ and $k$
 is to express both the intrinsic metric of $\Sigma$ and the extrinsic curvature in terms of 
the embedding functions $X^\mu(\sigma^1,\sigma^2)$, see for instance \cite{Polyakov:1986cs}.  The problem then reduces to a rather non-linear theory of the quantum fields $X^\mu(\sigma^1,\sigma^2)$. Although this direction
may deserve a further exploitation, in this paper we want to advocate a different route. We suggest to still consider the intrinsic metric on $\Sigma$ as the primary fluctuating degrees of freedom  and  treat  the extrinsic curvature of the surface as a  quantity  to a large extent derived from the intrinsic  metric. 
Below we give some arguments in favor of this point of view.

\subsection{Reconstructing  flat $R^3$ space from a 2d surface}

In this section  we closely follow the paper \cite{deBoer:2003vf}  in which a holographic reconstruction of Minkowski spacetime was proposed. In the present context,  the intrinsic metric on a closed  2d  surface is that 
holographic data which allows to reconstruct (with a certain  degree of uniqueness) the spacetime, the hypersurface $\Sigma$ and the extrinsic geometry of the surface. We stick to the picture where the entangling surface $\Sigma$ lies in a space-like 3d  hypersurface $R^3$ with flat geometry. The analysis below is focused  entirely on the flat space $R^3$.  

In a vicinity of surface $\Sigma$ one can always choose the normal coordinates  in which the metric  on  the space $R^3$ takes the form
\be
ds^2=dr^2+g_{ij}(r,x) dx^i dx^j\, ,
\lb{41}
\ee
where $r$ is a radial coordinate and  $x^i\, ,\, i=1,2$ are coordinates on $\Sigma$  defined as $r=0$, so that  $g_{ij}(r=0,x)=\gamma_{ij}(x)$ is the intrinsic  metric on $\Sigma$ .
The space $R^3$ is flat so that the Riemann tensor for metric (\ref{41}) is vanishing. This imposes the following conditions on $g_{ij}(r,x)$:
\be
g''=\frac{1}{2}g'g^{-1}g'\, ,
\lb{42-1}
\ee
\be
R_{likj}(g)=\frac{1}{4}g'_{ij}g'_{lk}-\frac{1}{4}g'_{ik}g'_{lj}\, ,
\lb{42-2}
\ee
\be
\nabla_k g'_{ij}-\nabla_j g'_{ik}=0\, ,
\lb{42-3}
\ee
where  $g'\equiv\partial_r g$.
Equation (\ref{42-1}) is equivalent to 
\be
g'''=0
\lb{42-4}
\ee
so that $g_{ij}(r,x)$ is quadratic in  the radial  coordinate  $r$
\be
g(r,x)=\gamma(x)+g^{(1)}(x)r+g^{(2)}(x)r^2\, .
\lb{43}
\ee
In fact, substituting decomposition  (\ref{43}) into equation (\ref{42-1}) we get that
\be
g^{(2)}=\frac{1}{4}g^{(1)} \gamma^{-1} g^{(1)}\,  .
\lb{44}
\ee
Furthermore, the equations (\ref{42-3}) and (\ref{42-2}) impose certain restrictions on the tensor $g^{(1)}_{ij}$:
\be
\nabla^j g^{(1)}_{ij}=\partial_i \tr (\gamma^{kl}g^{(1)}_{kl})
\lb{45}
\ee
and
\be
R(\gamma) &=& \frac{1}{4}\left((\tr(\gamma^{-1}g_{(1)}))^2-\tr(\gamma^{-1}g_{(1)})^2\right)\, \nonumber \\
&=&\frac{1}{2}\det (\gamma^{-1}g_{(1)})\, .
\lb{46}
\ee
A possible ``holographic'' interpretation of (\ref{45}) is as the  conservation law  of the stress-energy tensor $t_{ij}=g^{(1)}_{ij}-\gamma_{ij} \tr(\gamma^{-1}g_{(1)})$ while the  equation
(\ref{46}) imposes a  further restriction on $t_{ij}$. (One could say that   (\ref{46}) replaces the usual condition on the trace of a stress-energy tensor of a two-dimensional CFT by a non-linear one. Clearly, the corresponding field theory is not conformally invariant.)

The extrinsic curvature of surface $\Sigma$ is related to the tensor $g^{(1)}_{ij}$ as follows
\be
k_{ij}=\frac{1}{2}g^{(1)}_{ij}
\lb{47}
\ee
so that  eq.(\ref{46}) becomes the standard Gauss-Codazzi relation (the first equation in (\ref{40})).  Let us count the number of independent degrees of freedom contained in the extrinsic curvature. A priori, $k_{ij}$ has 3 components. The conservation law (\ref{45}) imposes 2 constraints and the equation (\ref{46}) gives one more constraint. Thus, the number of constraints is exactly equal to the number of components and there are no independent degrees of freedom left. 
We conclude that 
the tensor $g^{(1)}_{ij}$ (and, respectively, the extrinsic curvature (\ref{47})) is   determined by metric $\gamma$.   The space of parameters of the extrinsic curvature thus should be  finite dimensional. 
The functional integration over $k$ in (\ref{2}) then reduces to an ordinary finite dimensional integration over the modular parameters of $k$. 
In general the  analysis of the modular space of the extrinsic curvature may be rather complicated. Below we 
give an analysis of this problem  in the case of  the rotationally  invariant intrinsic  metrics.

\subsection{Rotationally invariant metrics}

We consider the case when the intrinsic metric on surface $\Sigma$ is rotationally invariant, i.e. it does not depend on one of the coordinates. It can be always brought to a conformally flat form, so that
\be
\gamma_{xx}=\gamma_{yy}=f(x)\, .
\lb{48}
\ee
For instance, if $\Sigma$ is the round sphere of radius $a$ then we have that $f(x)=a^2/\cosh^2(x)$ and the coordinate $y$ changes from $0$ to $2\pi$. In this coordinate system the stationary point of $O(2)$ symmetry corresponds to infinite values of $x$.

In what follows we assume that $y$ is periodic with period $2\pi$. Moreover, we assume that the stationary point of the rotational symmetry on the surface $\Sigma$ corresponds to the infinite values of $x$ where $f(x)$ vanishes and  $f'(x)/f(x)=\mp 2$ for $x=\pm\infty$, as in the case of the round sphere. 

Provided the 3d metric (\ref{41}) is rotationally invariant the solution to equations (\ref{45}), (\ref{46}) is 
\be
g^{(1)}_{yy}=\sqrt{\frac{Cf^2(x)-f'^2(x)}{f(x)}}\, , \ \ g^{(1)}_{xx}=\frac{2(f'^2(x)-f(x)f''(x))}{\sqrt{f(x)(Cf^2(x)-f'^2(x))}}\, .
\lb{49}
\ee
The constant $C$ is determined by the condition of regularity of the 3d metric at the stationary point of $O(2)$ symmetry, $C=f'^2/f^2|_{x=\infty}=4$.
(Note that if function $f(x)$ nowhere vanishes then $C$ remains an arbitrary integration  constant.)

The intrinsic curvature, calculated using relation (\ref{46})), 
\be
R_\Sigma=\frac{1}{2f^2}~g^{(1)}_{xx}g^{(1)}_{yy}=\frac{f'^2(x)-f(x)f''(x)}{f^3(x)}\, ,
\lb{50}
\ee
does not depend on constant $C$. The quantity $K_\Sigma$ defined  in the second equation in (\ref{40}),
\be
K_\Sigma &=&-\frac{1}{8f^2}(g^{(1)}_{xx}-g^{(1)}_{yy})^2\nonumber \\
&=&-\frac{1}{8f^3(x)(Cf^2(x)-f'^2(x))}(Cf^2(x)-3f'^2(x)+2f(x)f''(x))^2\, ,
\lb{51}
\ee
is negative for any 2d metric (\ref{48}). It vanishes if and only if  $f(x)=a^2/\cosh^2(x)$, i.e. when the surface $\Sigma$ is the round sphere.

\subsection{Surface instantons}

We consider a certain class of surfaces which  we call the {\it surface instantons}. These instantons are defined by the condition that components
(\ref{49})  are proportional to each other,
\be
g^{(1)}_{yy}=\lambda ~g^{(1)}_{xx}\, .
\lb{52}
\ee
The reason why we call them instantons is that for these surfaces the dynamical ``curvature''  $K_\Sigma$ is proportional to the ``topological'' curvature $R_\Sigma$ (similarly to the
instantons in  the Yang-Mills theory),
\be
K_\Sigma=-\frac{(1-\lambda)^2}{4\lambda}~  R_\Sigma \, .
\lb{53}
\ee
Let us fix the value of constant $C=4$. Then equation (\ref{52}) has a unique solution
 \be
 f_\lambda (x)=\frac{a^2}{\cosh^{2\lambda}(x/\lambda)}\, .
 \lb{54}
 \ee
 The corresponding intrinsic curvature is
 \be
 R_\Sigma (x)=\frac{2}{a^2}~\frac{1}{\lambda\cosh^{2(1-\lambda)}(x/\lambda)}\, .
\lb{55}
\ee
For negative $\lambda$ the curvature (\ref{55}) is negative and hence the 2d surface $\Sigma$  is hyperbolic.  For positive  $\lambda$ the curvature (\ref{55}) is positive and  the corresponding surfaces belong to the same topological class as the round sphere, the latter corresponds to value $\lambda=1$.  
The area of the surface is finite for any $\lambda>0$,
\be
A(\lambda)=2\pi^{3/2}a^2\frac{\Gamma (\lambda+1)}{\Gamma (\lambda+{1\over 2})}\, .
\lb{55-1}
\ee
If $\lambda >1$ the curvature (\ref{55}) is singular at $x=\pm\infty$. The singularity is, however, integrable.   In fact we find that
\be
\frac{1}{8\pi}\int_\Sigma R_\Sigma \sqrt{\gamma} =\theta(\lambda)\, ,
\lb{56}
\ee
where $\theta(\lambda)$ is the step function. Thus, the  surface instantons  with positive $\lambda$ are characterized by genus $h=0$ and they are  continuous deformations of the round sphere.
For $\lambda>0$ the quantity $\cal K$  is equal to
\be
{\cal K}=-\frac{(1-\lambda)^2}{4\lambda}\, .
\lb{57}
\ee
We observe that ${\cal K}$  is invariant under a duality  transformation $\lambda\rightarrow 1/\lambda$. Notice that this duality relates two  different geometries, as one can see from
(\ref{54}),  (\ref{55}).

The entanglement entropy of a surface instanton, as follows from eq.(\ref{30}), is
\be
S(\lambda)={NA(\lambda)\over 48\pi\epsilon^2}-{1\over 2}(A\pi^2+B\pi^2{(1-\lambda)^2\over 4\lambda})\ln \frac{A(\lambda)}{\epsilon^2}\, .
\lb{57-1}
\ee
For the round sphere $(\lambda=1$), as was observed in \cite{Solodukhin:2008dh} (see also \cite{Casini:2010kt} and \cite{Dowker:2010nq} for generalizations to spheres in higher dimensions),  the contribution of the anomaly of type B vanishes and the log term in the entropy is determined only by the anomaly of type A. However, as indicates eq.(\ref{57-1}), the anomaly of type B contributes to the entropy of a smoothly deformed  sphere (parametrized by $\lambda$).

\subsection{The averaging over surface instantons}
\lb{surf-inst}
In the mini-superspace approximation when the contribution of the surface instantons is taken into account the integration over the modular space of the extrinsic curvature reduces to 
 a finite-dimensional integral over parameter $\lambda$. The measure of integration
\be
{\cal D}\lambda=d\lambda ~\mu(\lambda)
\lb{58}
\ee
is naturally  to choose to be invariant under the duality transformation $\lambda\rightarrow 1/\lambda $. This imposes a restriction on the function $\mu(\lambda)$  in the measure,
\be
\mu(1/\lambda)=-\lambda^2 \mu(\lambda)\, .
\lb{59}
\ee
This equation fixes  $\mu(\lambda)$  up to a multiplication on a duality invariant function. Let us take this invariant function to be a power of $\lambda+1/\lambda$,
\be
\mu_\nu(\lambda)=\frac{\lambda^2-1}{\lambda^2}(\lambda+1/\lambda)^\nu\, .
\lb{60}
\ee
The power function may be is not the most general one, however, by using an inverse Mellin transform, any other possible function can be expressed in terms of power functions.

The subsequent procedure is as follows. We consider surfaces of spherical topology, i.e. with  genus $h=0$.
We first rescale  the intrinsic metric $\gamma \rightarrow \gamma/A$ so that the dependence on the area  $A$ is singled out.  
The surface action $W_\Sigma$ then rescales as
\be
W_\Sigma (\gamma)=W_\Sigma (\gamma/A(\Sigma))+(1-\alpha)\left(A\pi^2+B\pi^2\frac{(1-\lambda)^2}{4\lambda}\right)\frac{1}{2}\ln A(\Sigma)\, .
\lb{61}
\ee
The corresponding partition function factorizes
$$
Z(\alpha)=Z_A(\alpha)Z_B(\alpha)\, .
$$
The part that depends on the anomaly of type A is independent of $\lambda$, it contains only integration over the intrinsic metric $\gamma$ and   can be treated in the same way as in section 5. 
With the normalization defined in section 2 we obtain that
\be
Z_A(\alpha)= e^{-\frac{d(\alpha)A(\Sigma)}{8\pi\epsilon^2}} {\cal Z}_{d(\alpha)}(A(\Sigma))/{\cal Z}_{d=0}(A(\Sigma))\, ,
\lb{61-2}
\ee
where the effective number of scalar fields this time equals to  $d(\alpha)=-3A\pi^2(1-\alpha)$. Thus the contribution of the anomaly of the type A is the same as in section 5.

Let us now discuss  the part which is due to the anomaly of type B and which 
contains the integration over   $\lambda$. 
The averaging over the surface instantons then reduces to the integral
\be
Z_B(\alpha)=\int_0^\infty d\lambda \mu_\nu(\lambda) e^{-\beta (\lambda+1/\lambda-2)}\, ,
\lb{62}
\ee
where we introduced $\beta=\frac{B\pi^2}{8}(1-\alpha)\ln A(\Sigma)$. Suppose that $\nu=n$ is an integer. Then  the partition function (\ref{62}) is evaluated as follows
\be
Z_B(\alpha)=2\frac{n!}{\beta^{n+1}}\sum_{l=0}^n \frac{2^l}{l!}\beta^l\, .
\lb{62-1}
\ee 
In the calculation of  the corresponding contribution to the entropy we have to neglect a singular term which is due to
$\ln (1-\alpha)$ in $\ln Z_B(\alpha)$. This singular term does not depend on the area or any other geometric characteristics of the surface. Combining the contributions due to the anomalies of type A and B we arrive at the average entropy
\be
<S>=\frac{N\, A(\Sigma)}{48\pi\epsilon^2}-\frac{9}{10}A\pi^2\ln A(\Sigma)-(\nu+1)\ln(\ln  A(\Sigma)) +{1\over 4}B\pi^2\ln A(\Sigma)\, .
\lb{63}
\ee
The  two last terms are due to the anomaly of type B.
The last term comes from $l=1$ term in the sum (\ref{62-1}), it does not depend on $\nu$ (notice, however, that if $n=0$ this term does not appear in the sum (\ref{62-1})).
We notice the appearance in (\ref{63}) of  a rather unusual term $ \ln(\ln  A(\Sigma))$.  The exact coefficient in front of this term depends on the value of the parameter $\nu$ in the measure (\ref{60}). Its exact value can not be determined from the symmetry arguments which we used here. The entropy (\ref{63}) can now be analytically continued to any, not necessarily integer, value of $\nu$. It is an important open question whether the result (\ref{63}) remains unchanged in the  complete functional integral. We believe that the answer to this question is affirmative, however more work has to be done. 
Possibly, our result (\ref{63}) could be tested in  a numerical simulation of the entanglement entropy  in the spirit of the earlier works \cite{numericalworks}.

\section{ Conclusions}
\setcounter{equation}0

Entanglement entropy can be defined for any co-dimension two surface. The black hole horizon in this respect is not different from any other
surface of this type. The difference, however, appears when  the deformations of the entangling surface are considered. An a priori arbitrary surface can be
deformed in various ways without any influence on the  spacetime geometry  in which the surface is embedded. On the contrary, the horizon is 
rigidly embedded in the spacetime. So that by deforming the horizon one necessarily changes the spacetime geometry. That is why the entropy associated with a horizon
may manifest itself in the processes in which the horizon absorbs or emits. In each such act the spacetime geometry changes  thus allowing a distant observer to detect
the deformations happening at the horizon and to measure  the respective change in the entropy. This also explains why the fluctuations of the horizons are  rather different from 
the fluctuations of any other surface in the spacetime. The random changes in the geometry of the horizon are always accompanied by a corresponding random change in the spacetime
geometry, at least in a reasonably small vicinity of the horizon.
In this paper we have analyzed the  entropy of the random entangling surfaces in these two different cases provided the surface area  is kept fixed.
In all cases the analysis boils down to an effective two-dimensional problem which can be treated with the help of the known methods of the conformal field theory in two dimensions.
The main focus in the paper is made on the logarithmic corrections to the entropy.  We believe that our results may be useful both for the thermodynamics of horizons in  Quantum Gravity and in many applications to the condensed matter physics where the appearance of  random surfaces is a typical phenomenon.

\bigskip

The idea that the DDK results for random surfaces may be useful in the entanglement entropy calculations has originated  in the Summer of 1994 and has considerably evolved since then to  eventually take this present realization.
I am grateful to all people and circumstances which, directly or indirectly,  altogether have made this rather long project possible.

\bigskip

It is my pleasure to devote this paper to 75'th birthday of Stuart Dowker who has made a profound  and influential contribution to the modern quantum field theory.


\end{document}